\documentclass{aa}  
\usepackage{graphicx}
\usepackage[version=4]{mhchem}
\usepackage{units}
\usepackage{amsmath,amssymb}
\usepackage[utf8]{inputenc}
\usepackage[T1]{fontenc}
\usepackage{xcolor}
\usepackage{appendix}
\usepackage{booktabs}
\usepackage{multirow}
\DeclareUnicodeCharacter{2212}{-}
\usepackage{graphicx}
\usepackage{txfonts}
\begin{document}

   \title{Formation of carbonyl sulfide (OCS) via SH radicals in interstellar CO-rich ice under dense cloud conditions}


   \author{J. C. Santos
          \inst{1}
          \and
          H. Linnartz\inst{1, 2}
          \and
          K.-J. Chuang\inst{1}
          }


            \institute{Laboratory for Astrophysics, Leiden Observatory, Leiden University, PO Box 9513, 2300     RA Leiden, The Netherlands\\
                \email{santos@strw.leidenuniv.nl}
         \and Deceased, 31/12/2023
                }


 
  \abstract
   {Carbonyl sulfide (OCS) is widely observed in the gas phase towards star-forming regions and was the first of the only two sulfur-bearing species detected in interstellar ices so far. However, the chemical network governing its formation is still not fully understood. While the sulfurization of \ce{CO} and the oxidation of \ce{CS} are often invoked to form \ce{OCS}, other mechanisms could have a significant contribution. In particular, the multistep reaction involving \ce{CO} and \ce{SH} is a good candidate to forming \ce{OCS} in dense cloud environments.} 
   {We aim to constrain the viability of the $\ce{CO} + \ce{SH}$ route to forming solid \ce{OCS} in the interstellar medium, in a similar manner as $\ce{CO} + \ce{OH}$ is known to produce \ce{CO2} ice. This is achieved by conducting a systematic laboratory investigation of the targeted reactions on interstellar ice analogues under dense cloud conditions.}
   {An ultrahigh vacuum chamber is utilized to simultaneously deposit \ce{CO}, \ce{H2S}, and atomic \ce{H} at 10 K. \ce{SH} radicals produced in situ via hydrogen abstraction from \ce{H2S} react with \ce{CO} to form \ce{OCS}. The ice composition during deposition and subsequent warm-up is monitored by means of Fourier-transform reflection absorption infrared spectroscopy (RAIRS). Complementarily, desorbed species are recorded with a quadrupole mass spectrometer (QMS) during temperature-programmed desorption (TPD) experiments. Control experiments are performed to secure the product identification. The effect of different \ce{H2S}:\ce{CO} mixing ratios, with decreasing \ce{H2S} concentrations, on the \ce{OCS} formation yield is also explored.}
   {\ce{OCS} is efficiently formed through surface reactions involving \ce{CO}, \ce{H2S}, and \ce{H} atoms. The suggested underlying mechanism behind \ce{OCS} formation is $\ce{CO} + \ce{SH} \to \ce{HSCO}$ followed by $\ce{HSCO} + \ce{H} \to \ce{OCS} + \ce{H2}$. The \ce{OCS} yield reduces slowly, but remains significant with increasing \ce{CO}:\ce{H2S} mixing ratios (\ce{CO}:\ce{H2S} = 1:1, 5:1, 10:1, and 20:1).}
   {Our experiments provide unambiguous evidence that \ce{OCS} can be formed from $\ce{CO} + \ce{SH}$ in the presence of \ce{H} atoms. This route remains efficient for large \ce{H2S} dilutions ($5\%$ w.r.t \ce{CO}), suggesting that it is a viable mechanism in interstellar ices. Given that \ce{SH} radicals can be created in clouds throughout a wide evolutionary timescale, this mechanism could have a non-negligible contribution to forming interstellar \ce{OCS} ice.}

   \keywords{Astrochemistry, Methods: laboratory: solid state, Infrared: ISM, ISM: molecules}

   \maketitle
%

\section{Introduction}\label{sec:intro}

Among the over 300 molecules detected in the interstellar medium to date\footnote{https://cdms.astro.uni-koeln.de/}, those that contain sulfur constitute one of the most conspicuous chemical families. They are observed in the gas phase throughout a wide evolutionary time span, from clouds  (e.g., \citealt{Drdla1989, Navarro-Almaida2020, Spezzano2022, Esplugues2022}) to protostars (e.g., \citealt{Blake1987, Blake1994, vanderTak2003, Li2015, Drozdovskaya2018, Codella2021, delaVillarmois2023, Fontani2023, Kushwahaa2023}), protoplanetary disks \citep{Fuente2010, Phuong2018, Semenov2018, LeGal2019, Riviere-Marichalar2021, LeGal2021, Booth2024}, and solar-system bodies \citep{Smith1980, Bockelee-Morvan2000, Hibbitts2000, Jessup2007, Moullet2008, Moullet2013, Cartwright2020, Biver2021a, Biver2021b, Calmonte2016, Altwegg2022}. The hitherto detected species range in size, from simple diatomic molecules such as \ce{CS} and \ce{SO} to the thiols \ce{CH3SH} and \ce{CH3CH2SH} \citep{Linke1979, Gibb2000, Cernicharo2012, Koleniskova2014, Zapata2015, Muller2016, Majumdar2016, Rodriguez-Almeida2021}.

As opposed to gas-phase observations, however, only two sulfurated molecules have been detected in interstellar ices to date: \ce{OCS} and \ce{SO2} \citep{Palumbo1995, Boogert1997, Palumbo1997, Oberg2008, Boogert2022, McClure2023, Rocha2024}. This is probably in large part due to their low abundances combined with the intrinsic limitations of solid-state observations, such as the broadness of the features and their high degeneracy. Nonetheless, these confirmed ice species already bring vast chemical ramifications since they can act as reactants to form more complex organosulfur compounds, some of which with key biochemical roles (see, e.g., \citealt{Mcanally2024}). Yet, despite their astrochemical significance, many open questions still remain regarding the formation pathways for these small sulfur-bearing molecules in the solid-state.

One particularly important case is that of carbonyl sulfide (\ce{OCS}). It is a major gaseous sulfur carrier, commonly detected towards young stellar objects (e.g., \citealt{vanderTak2003, Herpin2009, Oya2016, Drozdovskaya2018, Kushwahaa2023, Lopez-Gallifa2024, Santos2024_submm}). It is also frequently detected in interstellar ices (see, e.g., \citealt{Palumbo1997, Boogert2022}), in part facilitated by the characteristically large absorption band strength of its 4.9 $\mu$m feature in comparison to other ice species (e.g., $\sim$1.2$\times$10$^{-16}$ cm molecule$^{-1}$ as measured by \citealt{Yarnall2022} for pure \ce{OCS} ice at 10 K; See also \citealt{Hudgins1993}). Observed \ce{OCS} abundances in both protostellar ices and hot-core gas point to a solid-state formation mechanism occurring predominately during the high-density pre-stellar core stage ($A_\text{V}>9$, $n_\text{H}\gtrsim10^5$ cm$^{-3}$), after the onset of the \ce{CO} catastrophic freeze-out \citep{Palumbo1997, Boogert2022, Santos2024_submm}. This proposed icy origin of \ce{OCS} is corroborated by gas-phase astrochemical models, which fail to reproduce observed \ce{OCS} abundances \citep{Loison2012}. In the inner regions of the protostellar envelope, thermal heating by the protostar causes the volatile ice content to fully sublimate, enabling the bulk of the gaseous \ce{OCS} observations.

So far, the two most commonly proposed routes to \ce{OCS} involve either the oxidation of \ce{CS} or the sulfurization of \ce{CO} in the solid state \citep{Palumbo1997, Ferrante2008, Adriaens2010, Chen2015, Laas2019, Shingledecker2020, Boogert2022}:

\begin{equation}
    \ce{CS} + \ce{O} \to \ce{OCS},
    \label{eq:CS+O}
\end{equation}

\begin{equation}
    \ce{CO} + \ce{S} \to \ce{OCS}.
    \label{eq:CO+S}
\end{equation}

\noindent However, \ce{CS} abundances are considerably smaller ($\lesssim2.5\%$) than those of \ce{OCS}, casting doubts on the extent of the contribution from Reaction \ref{eq:CS+O} to the interstellar \ce{OCS} content \citep{Boogert2022}. Reaction \ref{eq:CO+S} is more often invoked as the dominant route to \ce{OCS}, especially due to the large availability of solid-state \ce{CO} as a reactant in the post freeze-out stage.

Hydrogen in its atomic form is also abundant in pre-stellar cores (\ce{H}/\ce{CO}$\sim$1$-$10, \citealt{Lacy1994, Goldsmith2005}) and can trigger solid-state reactions of relevance to the sulfur network. Following adsorption onto dust grains, sulfur atoms are subject to successive hydrogenation reactions to form \ce{SH} and \ce{H2S} via the steps:

\begin{equation}
    \ce{S} + \ce{H} \to \ce{SH},
    \label{eq:S+H}
\end{equation}

\begin{equation}
    \ce{SH} + \ce{H} \to \ce{H2S},
    \label{eq:SH+H}
\end{equation}

\noindent both of which are predicted by astrochemical models to proceed very efficiently (see, e.g., \citealt{Garrod2007, Druard2012, Esplugues2014, Vidal2017}). Once \ce{H2S} ice is formed, it can further react with another hydrogen atom via an abstraction route \citep{Lamberts2017, Oba2018, Oba2019, Santos2023}, or be dissociated through energetic processing (e.g., \citealt{Jimenez-Escobar2014, Chen2015, Cazaux2022}):
\begin{subequations}
\begin{equation}
    \ce{H2S} + \ce{H} \to \ce{SH} + \ce{H2},
    \label{eq:H2S+H}
\end{equation}

\begin{equation}
    \ce{H2S} + h\nu \to \ce{SH} + \ce{H},
    \label{eq:H2S+hv}
\end{equation}
\end{subequations}

\noindent thus partially replenishing the \ce{SH} ice content. As a result, \ce{SH} radicals will likely be present in the ice throughout a wide evolutionary time span, and could potentially serve as an alternative source of sulfur to forming \ce{OCS}.

A particularly promising pathway is the reaction between \ce{SH} and \ce{CO} followed by a hydrogen abstraction step \citep{Adriaens2010, Chen2015}:

\begin{equation}
    \ce{SH} + \ce{CO} \to \ce{HSCO},
    \label{eq:SH+CO}
\end{equation}

\begin{equation}
    \ce{HSCO} + \ce{H} \to \ce{OCS} + \ce{H2}.
    \label{eq:HSCO+H}
\end{equation}

\noindent which involves the formation of the intermediate complex \ce{HSCO}. This pathway is analogous to the case of \ce{CO2} ice formation from \ce{CO} and \ce{OH} through the \ce{HOCO} complex (e.g., \citealt{Oba2010a, Ioppolo2011, Noble2011, Qasim2019, Molpeceres2023}). In the case of \ce{CO2}, \cite{Molpeceres2023} find that the spontaneous dissociation of \ce{HOCO} into \ce{H} and \ce{CO2} is not energetically viable, and thus the formation of the latter is only possible through the interaction of the \ce{HOCO} intermediate with a hydrogen atom. Similarly, \ce{HSCO} is also prevented from spontaneously dissociating into \ce{OCS} and \ce{H} \citep{Adriaens2010}. In the laboratory, \ce{OCS} is readily formed in energetically-processed \ce{CO}:\ce{H2S} ice mixtures \citep{Ferrante2008, Garozzo2010, Jimenez-Escobar2014, Chen2015}, for which both reactions involving $\ce{CO} + \ce{S}$ and $\ce{CO} + \ce{SH}$ have been suggested as possible formation routes. Moreover, \cite{Nguyen2021} allude to the possibility of forming \ce{OCS} through reactions \ref{eq:SH+CO} and \ref{eq:HSCO+H} in a \ce{CO}:\ce{H2S} ice mixture exposed to hydrogen atoms, albeit without exploring it further. Dedicated laboratory studies focused on assessing this particular reaction pathway to \ce{OCS} formation are therefore still highly warranted.

In this experimental work, we investigate the viability of \ce{OCS} formation via Reactions \ref{eq:SH+CO} and \ref{eq:HSCO+H} under dark cloud conditions by depositing simultaneously \ce{H2S}, \ce{CO}, and \ce{H} atoms at 10 K. The \ce{SH} radicals are produced via Reaction \ref{eq:H2S+H}, and are subsequently subject to reacting with neighboring species. The experimental methods utilized in this work are described in Section \ref{sec:methods}. In Section \ref{sec:results_diss}, we present and discuss our main results. Their astrophysical implications are elaborated in Section \ref{sec:astro}, followed by a summary of our main findings in Section \ref{sec:conc}.


\section{Experimental methods}\label{sec:methods}

\begin{table*}[htb!]
\centering
\caption{Experiments performed in this work.}
\label{tab:exp_list} 
\begin{tabular}{lccccc}  
\toprule\midrule
Experiment                              &   Label  &   \ce{CO} flux            &   \ce{H2S} flux               &   \ce{H} flux             &    \ce{CO}:\ce{H2S}:\ce{H}\\
                                        &          &   (cm$^{-2}$ s$^{-1}$)    &   (cm$^{-2}$ s$^{-1}$)        &   (cm$^{-2}$ s$^{-1}$)    &\\                                               
\midrule
$\ce{CO} + \ce{H2S} + \ce{H}$           &   1      &   $1.8\times10^{11}$      &   $1.8\times10^{11}$          &   $5.5\times10^{12}$      &    1:1:30\\
$\ce{CO} + \ce{H2S}$                    &   2      &   $1.8\times10^{11}$      &   $1.8\times10^{11}$          &   $-$                     &    1:1:0\\
$\ce{^{13}C^{18}O} + \ce{H2S} + \ce{H}$ &   3      &   $1.8\times10^{11}$      &   $1.8\times10^{11}$          &   $5.5\times10^{12}$      &    1:1:30\\
$\ce{CO} + \ce{H2S} + \ce{H}$           &   4      &   $9.3\times10^{11}$      &   $1.8\times10^{11}$          &   $5.5\times10^{12}$      &    5:1:30\\
$\ce{CO} + \ce{H2S} + \ce{H}$           &   5      &   $1.8\times10^{12}$      &   $1.8\times10^{11}$          &   $5.5\times10^{12}$      &    10:1:30\\
$\ce{CO} + \ce{H2S} + \ce{H}$           &   6      &   $3.7\times10^{12}$      &   $1.8\times10^{11}$          &   $5.5\times10^{12}$      &    20:1:30\\
\midrule\bottomrule
\end{tabular}
\end{table*}

This work utilizes the ultrahigh vacuum (UHV) setup SURFRESIDE$^3$, for which detailed descriptions are provided elsewhere \citep{Ioppolo2013, Qasim2020}. In a nutshell, this setup contains a main chamber that operates at base pressures of  $\sim5\times10^{-10}$ mbar, and at the center of which a gold-plated copper substrate is mounted on the tip of a closed-cycle helium cryostat. Resistive heaters are used to vary the substrate temperature between 8 and 450 K, as monitored by two silicon diode sensors with a relative accuracy of 0.5 K. A hydrogen atom beam source (HABS \citealt{Tschersich2000}) generates \ce{H} atoms that are subsequently cooled down by colliding with a nose-shaped quartz pipe before reaching the substrate. Simultaneously to in-letting \ce{H} atoms into the chamber, gases of \ce{H2S} (Linde, 99.5\% purity) and \ce{CO} (Linde, 99.997\% purity) are concomitantly admitted through two separate all-metal leak valves. The heavier isotopologue \ce{^{13}C^{18}O} (Sigma-Aldrich, 95\% purity \ce{^{18}O}, 99\% purity \ce{^{13}C}) is also utilized to assist in the product identification. Fourier-transform reflection-absorption infrared spectroscopy (FT-RAIRS) is used to monitor ice growth in the range of 700 to 4000 cm$^{-1}$ with 1 cm$^{-1}$ spectral resolution. When the deposition is complete, temperature-programmed desorption experiments (TPD) are performed by heating the sample at a constant rate of 5 K min$^{-1}$. During TPD, the solid phase is monitored by RAIRS, while the sublimated species are immediately ionized by 70 eV electron impact and are recorded by a quadrupole mass spectrometer (QMS). The variation of the substrate temperature during the collection of each IR TPD spectrum is of $\sim$10 K.

The absolute abundance of the ice species can be derived from their integrated infrared absorbance ($\int Abs(\nu)d\nu$) using a modified Beer-Lambert law:

\begin{equation}
    N_X=\ln10\frac{\int Abs(\nu)d\nu}{A'(X)}
    \label{eq:N_RAIRS}
\end{equation}

\noindent where $N_X$ is the column density in molecules cm$^{-2}$ and $A'(X)$ is the apparent absorption band strength in cm molecule$^{-1}$ of a given species. We utilize $A'(\ce{CO})_{\sim2142\text{cm}^{-1}}\sim(4.2\pm0.3)\times10^{-17}$ cm molecule$^{-1}$ and $A'(\ce{H2S})_{\sim2553\text{cm}^{-1}}\sim(4.7\pm0.1)\times10^{-17}$ cm molecule$^{-1}$, as measured for our reflection-mode IR settings using the laser-interference technique (see \citealt{Santos2023} for the case of \ce{H2S}). For the target product, \ce{OCS}, we employ the band strength of $A'(\ce{OCS})_{\sim2043\text{cm}^{-1}}\sim(1.2\pm0.1)\times10^{-16}$ cm molecule$^{-1}$ measured in transmission mode by \cite{Yarnall2022}, and correct it by an averaged transmission-to-reflection conversion factor of 3.2 derived with our experimental setup (see \citealt{Santos2023} for the case of \ce{H2S}, which was later combined with \ce{CO} ice depositions to derive the averaged setup-specific conversion factor).

Table \ref{tab:exp_list} summarizes the experiments performed in this work. All depositions are carried out for 180 minutes with a constant substrate temperature of 10 K. The relative errors of the \ce{H}-atom flux, as well as both molecule fluxes, are estimated to be  $\lesssim5\%$. The latter are estimated from evaluating the ice growth rate in multiple pure ice deposition experiments. To estimate the former, multiple experiments to calibrate the relative H-atom flux are conducted. These consist of trapping \ce{H} atoms inside \ce{O2} ice matrices and monitoring the formation of \ce{HO2}, which is proportional to the \ce{H}-atom flux provided that \ce{O2} is overabundant relative to \ce{H} \citep{Ioppolo2013, Fedoseev2015}. The instrumental uncertainties in the integrated QMS and IR signals are derived from the corresponding integrated spectral noise for the same band width.

\section{Results and discussion}\label{sec:results_diss}
\subsection{OCS formation}

\begin{figure*}[htb!]\centering
\includegraphics[scale=0.5]{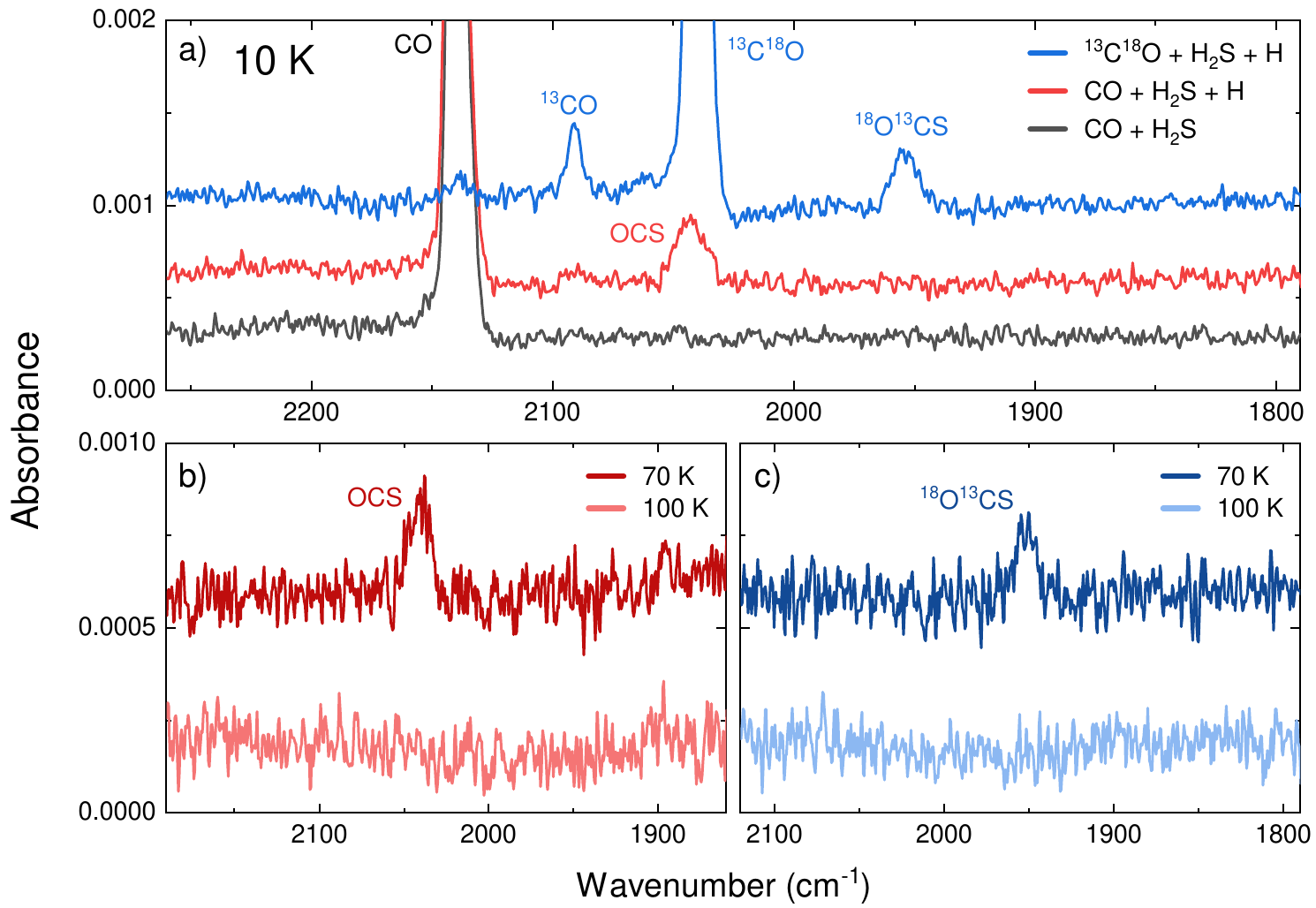}
\caption{Infrared data that confirms the formation of \ce{OCS} from \ce{CO}, \ce{H2S}, and \ce{H} atoms. Panel a): Spectrum collected after deposition of $\ce{CO} + \ce{H2S} + \ce{H}$ (experiment 1, red) together with the control experiment without \ce{H} atoms (experiment 2, grey) and the isotope-labeled experiment with $\ce{^{13}C^{18}O}$ (experiment 3, blue). Small features assigned to \ce{^{13}CO} and \ce{^{12}CO} are due to minor contaminations from the gas bottle (respectively, $\sim4\%$ and $\sim2\%$ with respect to \ce{^{13}C^{18}O}). Panel b): Infrared spectra acquired during TPD following the deposition of experiment 1. The spectrum in dark red (upper) is taken at $\sim$70 K, and the one in light red (lower) is taken at $\sim$100 K. Panel c): Same as panel b), but for the isotope-labeled experiment. In all panels only the relevant frequency range is shown and the spectra are offset for clarity.}
\label{fig:IR_spectra}
\end{figure*}

Panel a) of Figure \ref{fig:IR_spectra} shows the infrared spectra obtained after codeposition of \ce{CO}, \ce{H2S}, and \ce{H} atoms (1:1:30, experiment 1) at 10 K, as well as after a control experiment of only \ce{CO} and \ce{H2S} (1:1, experiment 2). When \ce{CO} and \ce{H2S} are exposed to hydrogen atoms, a new feature appears at $\sim2043$ cm$^{-1}$ ($\sim$4.89 $\mu$m) with a full-width at half maximum (FWHM) of $\sim$15 cm$^{-1}$ ($\sim$0.03 $\mu$m). Based on its peak position and the ice elemental composition (i.e., C, O, S, and H), this feature is assigned as the \ce{CO}-stretching mode of \ce{OCS} ($\nu_1$, \citealt{Yarnall2022})\footnote{The numbers assigned to the \ce{CS} and \ce{CO} stretches of \ce{OCS} are interchangeable. In this work, we adopt the notation from \cite{Yarnall2022}.}. The band's FWHM and peak position are also consistent with previously reported values for \ce{OCS} in \ce{CO}-rich ices measured in reflection mode \citep{Ferrante2008}. To confirm this assignment, an analogous experiment is performed with a heavier isotopologue of carbon monoxide, \ce{^{13}C^{18}O} (experiment 3, blue spectrum in Figure \ref{fig:IR_spectra}). In this case, the \ce{^{18}O^{13}CS} $\nu_1$ feature appears at $\sim1954$ cm$^{-1}$ ($\sim$5.12 $\mu$m) in accordance with the expected redshift of the heavier isotopologue. During TPD, both features at $\sim2043$ cm$^{-1}$ and $\sim1954$ cm$^{-1}$ disappear from the ice in the same temperature interval of $70-100$ K (Panels b) and c) in Figure \ref{fig:IR_spectra}), thus signaling that the two bands correspond to the same molecule.

\begin{figure*}[htb!]\centering
\includegraphics[scale=0.5]{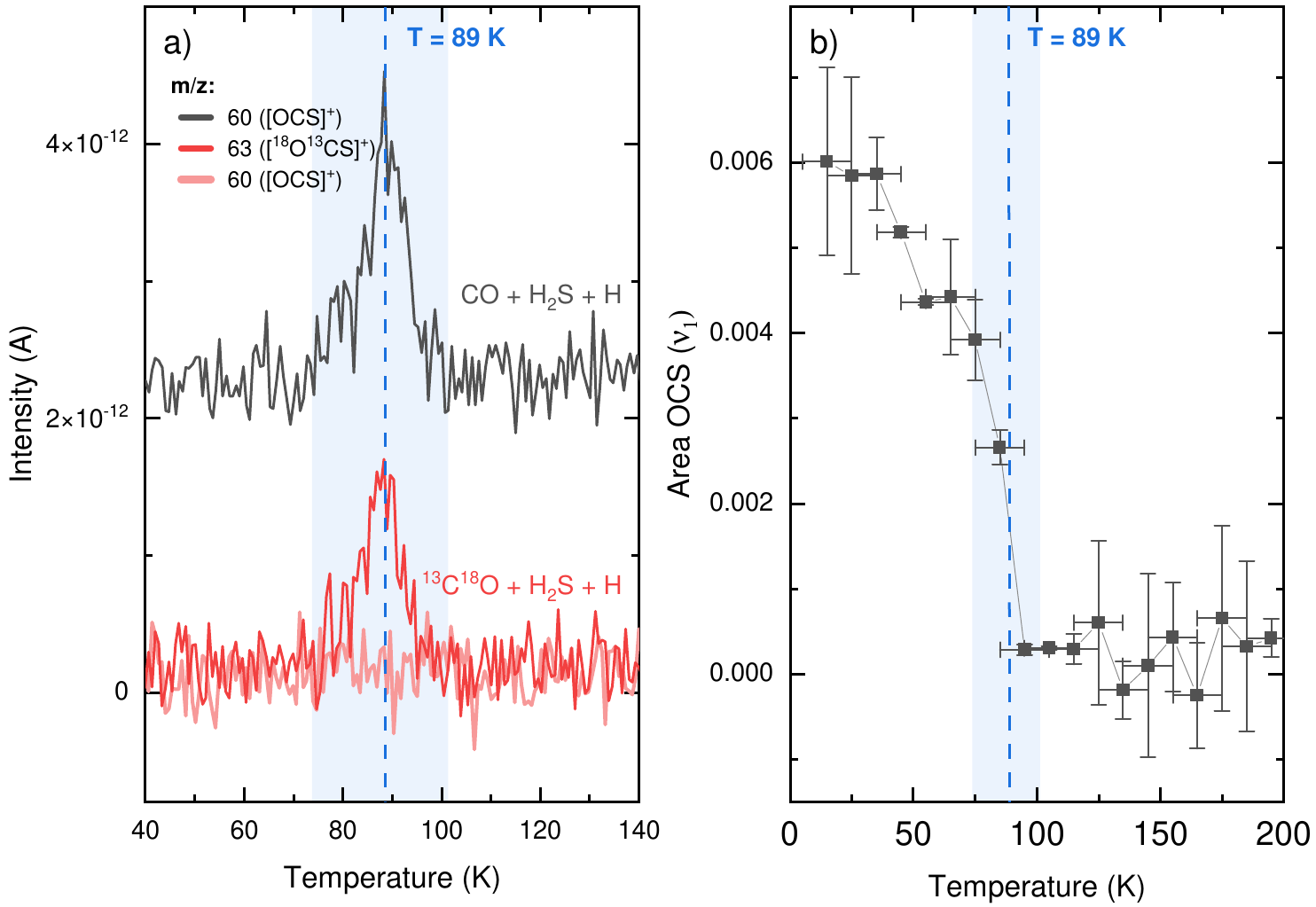}
\caption{Complementary QMS and infrared data acquired during TPD that confirms the \ce{OCS} detection. Panel a): TPD-QMS signal recorded for m/z = 60 after the deposition of $\ce{CO} + \ce{H2S} + \ce{H}$ (experiment 1, grey), as well as those for m/z = 63 (dark red) and m/z = 60 (light red) after the deposition of $\ce{^{13}C^{18}O} + \ce{H2S} + \ce{H}$ (experiment 3). The blue shadowed region denotes the range of desorption, and the dashed line highlights the peak desorption temperature. Signals from different experiments are offset for clarity. Panel b): Area of the \ce{OCS} $\nu_1$ band as a function of temperature during TPD following the deposition of experiment 1. The uncertainties in the horizontal axis are due to the substrate temperature variation of $\sim$10 K during the collection of each IR TPD spectrum. The blue shadowed area and dashed line are reproductions of the range and peak position shown in Panel a).}
\label{fig:QMS}
\end{figure*}

Complementarily to the infrared spectra, data obtained by the QMS during TPD can also be used to strengthen the \ce{OCS} identification. Panel a) of Figure \ref{fig:QMS} displays the signal for m/z = 60 recorded during the warm-up of the ice sample in the \ce{CO} + \ce{H2S} + \ce{H} experiment. This mass-to-charge ratio corresponds to the molecular ion of \ce{OCS}, its strongest signal resulting from 70 eV electron impact as listed in NIST\footnote{https://webbook.nist.gov/chemistry/}. A desorption feature appears peaking at $\sim$89 K, in agreement with previously reported desorption temperatures of OCS \citep{Ferrante2008, Nguyen2021}. In the isotope-labeled experiment, a similar desorption peak is observed for m/z = 63, consistent with the mass shift corresponding to the molecular ion of \ce{^{18}O^{13}CS}. Moreover, no peak signal is detected for m/z = 60, indicating that this same feature appearing in experiment 1 corresponds uniquely to carbonyl sulfide. The desorption temperature of \ce{OCS} as measured by the QMS correlates with the disappearance of the $\sim2043$ cm$^{-1}$ feature in the infrared spectra taken during TPD, as evinced by the area of this infrared band as a function of substrate temperature (Figure \ref{fig:QMS} panel b). Both the infrared and QMS data therefore provide unequivocal evidence for the formation of \ce{OCS} as a result of interactions between \ce{H2S}, \ce{CO}, and \ce{H} atoms in the solid state.

Aside from \ce{OCS}, another \ce{S}-bearing species that could presumably be formed under our experimental conditions is thioformic acid, \ce{HCOSH}. Indeed, formation routes via either the hydrogenation of the \ce{HSCO} intermediate or the interaction between $\ce{HCO} + \ce{SH}$ have been proposed by \cite{Rodriguez-Almeida2021} to explain \ce{HCOSH} observations towards the giant molecular cloud G+0.693–0.027, the former of which was later verified theoretically by \cite{Molpeceres2021}. Although \ce{HCOSH} has been shown to form upon keV electron irradiation of \ce{H2S}:\ce{CO} ices \citep{Wang2022}, \cite{Nguyen2021} only tentatively identify this product in similar hydrogenation experiments on \ce{H2S}:\ce{CO} ices. No evidence for this species is observed in our experiments above the instrumental detection limit. Likewise, we do not detect any signal of volatile sulfur allotropes such as \ce{S2} or \ce{S3}, nor of hydrogenated sulfur chains such as \ce{H2S2} or \ce{H2S3}---which could presumably be formed from the association of sulfur atoms produced via hydrogen abstraction reactions from \ce{SH} radicals. This leads to the conclusion that the potential production of atomic \ce{S} does not proceed efficiently under our experimental conditions. Moreover, since no \ce{H2S2} is detected, the recombination of two \ce{SH} radicals is also likely not efficient in the present experiments---signaling that \ce{HS} radicals are efficiently consumed, both by reacting with \ce{CO} leading to \ce{OCS} and by reacting with \ce{H} to reform \ce{H2S}.

Reactions \ref{eq:SH+CO} and \ref{eq:HSCO+H} are therefore the predominant contributors to forming \ce{OCS} in our experiments. As shown by \cite{Nguyen2021}, however, the backward reactions are also viable. Once formed, \ce{OCS} can be hydrogenated into the complex intermediate \ce{HSCO}, which in turn can further react with another \ce{H} atom to yield \ce{CO} and \ce{H2S}, as well as dissociate back into the reactants \ce{SH} and \ce{CO} \citep{Adriaens2010, Nguyen2021, Molpeceres2021}. The effective amount of \ce{OCS} detected will, therefore, be subject to these two competing directions.

\subsection{Effects of larger CO fractions}

\begin{figure}[htb!]\centering
\includegraphics[scale=0.5]{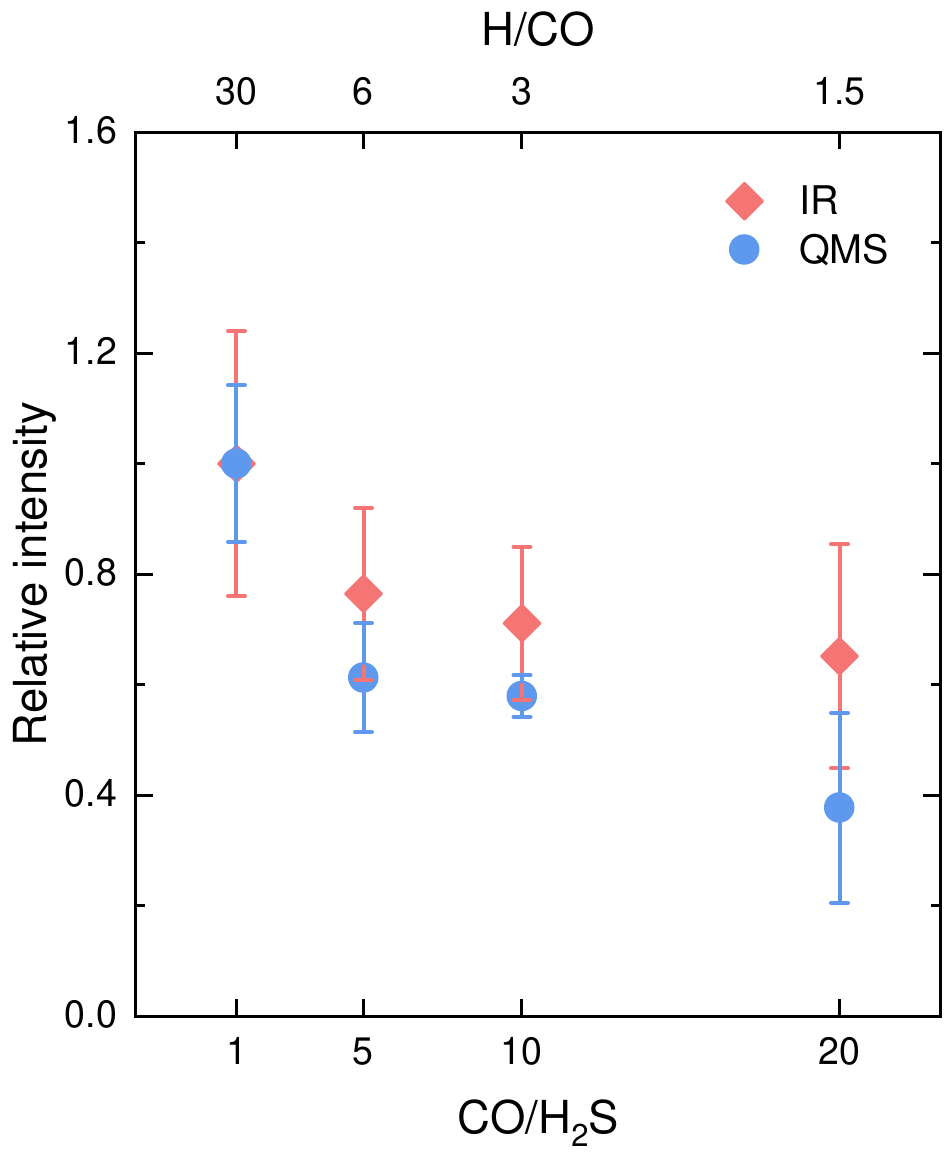}
\caption{Integrated intensities of the infrared and QMS features of \ce{OCS} for different flux conditions: \ce{CO}:\ce{H2S} = 1:1, 5:1, 10:1, and 20:1. The infrared value is derived from the area of the $\nu_1$ mode of \ce{OCS}, while the QMS counterpart is calculated from the integrated desorption band for m/z = 60 peaking at $\sim$89 K. The resulting yields are normalized to that of experiment 1 (\ce{CO}:\ce{H2S} = 1:1). The upper x-axis shows the corresponding ratios of \ce{H}/\ce{CO}, signaling the decrease in \ce{H}-atom availability with increasing \ce{CO} fractions.}
\label{fig:NOCS_ratios}
\end{figure}

Once the possibility of forming \ce{OCS} from \ce{CO} and \ce{SH} is ascertained, the next question that arises is regarding the viability of this route in more astronomically representative ice mixing ratios. To date, there are no convincing detections of \ce{H2S} in interstellar ices, with upper limits estimated to be $\leq$0.7\% with respect to water \citep{JimenezEscobar2011}. This would translate to ice abundance upper limits of roughly a few percent with respect to \ce{CO} (e.g., $N$(\ce{H2S})/$N$(\ce{CO}) $\lesssim0.035$ if assuming $N$(\ce{CO})/$N$(\ce{H2O}) $\sim0.21$ according to the median \ce{CO} ice abundance value w.r.t \ce{H2O} derived for low-mass young stellar objects (LYSOs) in \citealt{Boogert2015}). Accordingly, similar depositions with larger fractions of \ce{CO} with respect to \ce{H2S} are performed to assess how the dilution of the latter affects the yields of \ce{OCS}. The deposition fluxes of \ce{H2S} and atomic \ce{H} are kept the same in all experiments, with only variations in the \ce{CO} flux (see Table \ref{tab:exp_list}). Figure \ref{fig:NOCS_ratios} compares the relative intensities of the \ce{OCS} signals obtained from both the IR and QMS data for the different mixing ratios explored here (i.e., \ce{CO}:\ce{H2S} = 1:1, 5:1, 10:1, and 20:1). The areas of the \ce{OCS} $\nu_1$ vibrational modes for each mixing ratio after deposition are normalized to that of experiment 1, which yielded the largest absolute amount of products. Similarly, the areas of the m/z = 60 desorption band at $\sim89$ K are also normalized to that of experiment 1.

Overall, the infrared and QMS data provide consistent \ce{OCS} yields within their respective error bars, and signal that dilution of \ce{H2S} in \ce{CO} reduces, but does not impede \ce{OCS} formation under our experimental conditions. Indeed, in the highest dilution case (5$\%$ \ce{H2S} w.r.t. \ce{CO}), \ce{OCS} production diminishes by $\sim50\%$ (the mean value between the infrared and QMS results) in comparison to the non-diluted reference experiment (100$\%$ \ce{H2S} w.r.t. \ce{CO}). This decrease in \ce{OCS} yield is likely governed by the reduced number of available \ce{H} atoms in experiments with larger \ce{CO} fractions, as opposed to a diminishing effectiveness of the \ce{OCS} formation. As well established by previous works, the interaction of \ce{CO} with \ce{H} atoms in the solid state initiates a chain of reactions leading to \ce{H2CO} and \ce{CH3OH} \citep{Tielens1982, Charnley1992, Hiraoka1994, Watanabe2002, Fuchs2009, Santos2022}, which efficiently consumes the available atomic hydrogen in the system. Increasing the \ce{CO} deposition flux while maintaining a fixed \ce{H}-atom flux, therefore, results in less hydrogen being able to react with other species. This decrease in available \ce{H} atoms will impact more significantly the formation of \ce{SH} radicals (Reaction \ref{eq:H2S+H}) than the final abstraction step (Reaction \ref{eq:HSCO+H}), since the latter is predicted to proceed barrierlessly \citep{Adriaens2010}, whereas the former requires quantum tunneling through a barrier of $\sim$1500 K \citep{Lamberts2017}. The slowly decreasing trend in Figure \ref{fig:NOCS_ratios} suggests that \ce{OCS} can still be formed at 10 K in ice mixtures with \ce{CO}:\ce{H2S} higher than 20:1 and \ce{H}:\ce{CO} as low as 1.5:1. The formation of \ce{OCS} from $\ce{CO} + \ce{SH}$ could therefore have a non-negligible contribution to the observed \ce{OCS} in interstellar ices, as discussed below.

\section{Astrophysical implications}\label{sec:astro}

\begin{figure}[htb!]\centering
\includegraphics[scale=0.53]{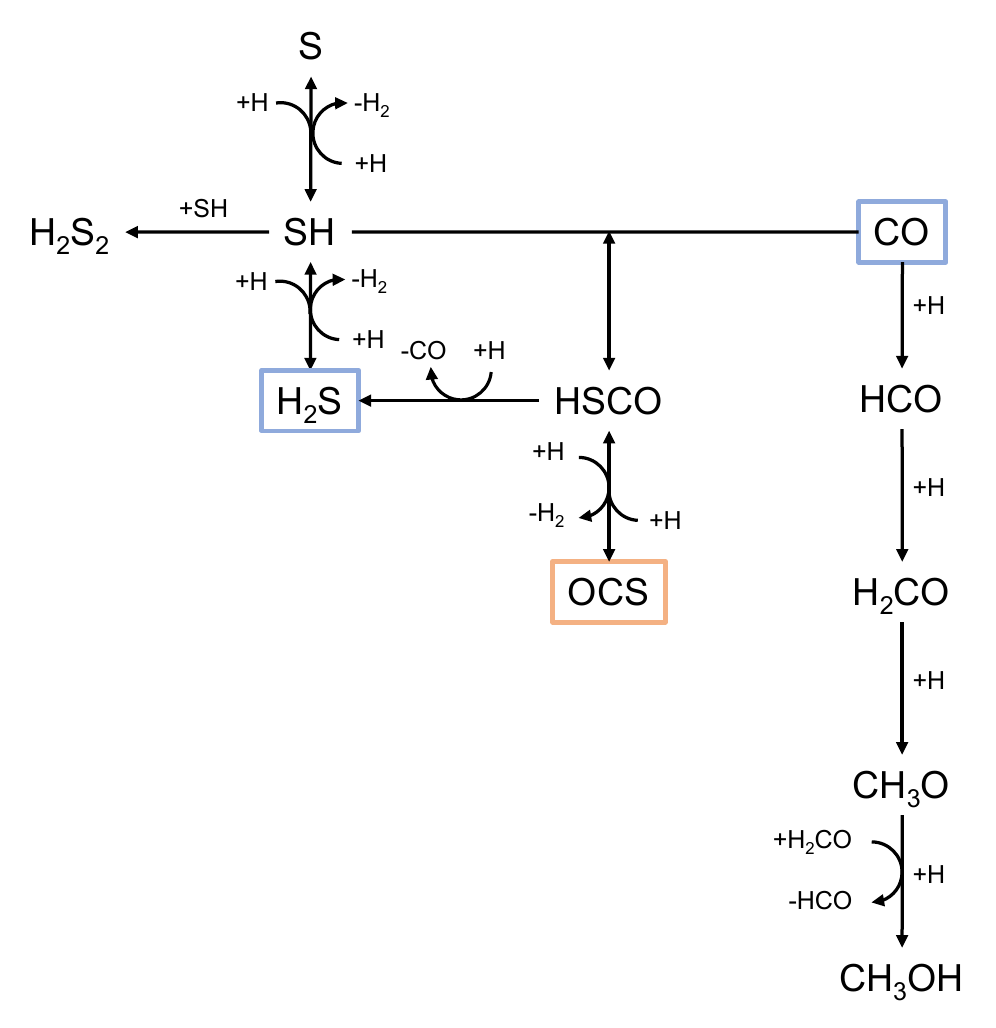}
\caption{Reaction scheme proposed in this work. Blue boxes denote the deposited molecules and the orange box highlights the studied product, \ce{OCS}. Hydrogren abstraction reactions from \ce{CH3OH}, \ce{H2CO}, and associated radicals are omitted for the sake of simplicity.}
\label{fig:network}
\end{figure}

The reaction network probed in this work is shown in Figure \ref{fig:network}. Aside from \ce{OCS}, other closed-shell species also formed within this network are \ce{H2S2} \citep{Santos2023}, \ce{H2CO} and \ce{CH3OH} \citep{Tielens1982, Charnley1992, Hiraoka1994, Watanabe2002, Fuchs2009, Santos2022}. In our experiments, \ce{H2S} is utilized as a source of \ce{SH} radicals via a hydrogen abstraction (Reaction \ref{eq:H2S+H}) to avoid reactions with \ce{S} atoms to interfere with the analysis. In the interstellar medium, however, \ce{SH} will not only be produced from \ce{H2S}, but also from the hydrogenation of sulfur atoms that adsorb onto dust grains (Reaction \ref{eq:S+H}). Observations of singly- and doubly-deuterated \ce{H2S} in Class 0 sources suggest that the bulk of its formation takes place early in the evolution of a cloud, before the catastrophic CO freeze-out stage begins \citep{Ceccarelli2014}. Accordingly, \ce{SH} radicals will be available in the ice as early as during the low-density pre-stellar core stage. As the cloud evolves and the density of the environment increases, most of the available \ce{S} atoms will be converted into \ce{H2S}. Nonetheless, \ce{SH} radicals can still be formed through Reaction \ref{eq:H2S+H} or through dissociation reactions induced by energetic processing. As a result, \ce{SH} will remain a viable reactant throughout a large fraction of a cloud's lifetime. 

Observations of \ce{OCS} towards massive young stellar objects (MYSOs) reveal that its 4.9 $\mu$m feature is best fitted by reference spectra of \ce{OCS} in a \ce{CH3OH}-rich environment \citep{Palumbo1997, Boogert2022}. Their column density ratios ($N$(\ce{OCS})/$N$(\ce{CH3OH})) in both ice and gas have similarly narrow distributions and comparable median values (within a factor of 3, \citealt{Santos2024_submm}), strengthening the hypothesis that both coexist in similar ice environments. Moreover, observed ice abundances of \ce{OCS} and \ce{CH3OH} are correlated \citep{Boogert2022}, suggesting a strong chemical link between the two molecules. Since methanol is known to be formed via the hydrogenation of \ce{CO} in the solid phase, such observations are in line with \ce{CO} being a common precursor between the two. This would imply that the bulk of \ce{OCS} ice likely originates deep into dense clouds, in a dense environment post catastrophic \ce{CO} freeze-out. Conversely, a large portion of the free sulfur atoms will likely be readily converted into \ce{H2S} at such densities. The reaction route proposed here could partially circumvent this issue since \ce{SH} radicals can be efficiently produced through a top-down mechanism (Reaction \ref{eq:H2S+H}). Moreover, estimated \ce{H2S} upper limits in ices are larger than the detected \ce{OCS} abundances \citep{Palumbo1997, JimenezEscobar2011, Boogert2022}, and thus this route remains viable in light of the observational evidence available so far. Overall, both $\ce{CO} + \ce{S}$ and $\ce{CO} + \ce{SH}$ routes are likely to be contributing to the observed \ce{OCS} abundances in interstellar ices, the extent to which warrants dedicated chemical modeling to be constrained.

\section{Conclusions}\label{sec:conc}

In this work, we conduct a systematic experimental investigation on the viability of forming \ce{OCS} under dark cloud conditions from ice mixtures of \ce{CO}, \ce{H2S}, and \ce{H} atoms. Our main findings are summarized below:

\begin{itemize}
    \item \ce{OCS} can be efficiently formed at 10 K from solid-state reactions involving \ce{CO}, \ce{H2S} and \ce{H}. The proposed underlying mechanism is via $\ce{CO} + \ce{SH} \to \ce{HSCO}$ followed by $\ce{HSCO} + \ce{H} \to \ce{OCS} + \ce{H2}$, analogously to \ce{CO2} ice.
    \item The \ce{OCS} product yield decreases slowly with increasing initial \ce{CO}/\ce{H2S} ratios and decreasing \ce{H}-atom availability, resulting in $\sim50\%$ less \ce{OCS} formation for a 20 times higher \ce{CO} abundance relative to \ce{H2S} and fixed \ce{H}-atom fluxes. This trend suggests that \ce{OCS} can be efficiently formed via the proposed route in more representative interstellar ices where \ce{H2S} is likely highly diluted in \ce{CO}.
    \item \ce{SH} can be produced both through bottom-up ($\ce{S} + \ce{H} \to \ce{SH}$) and top-down ($\ce{H2S} + \ce{H} \to \ce{SH} + \ce{H2}$) routes, thus making it a favorable reactant to form \ce{OCS} during the high-density, post \ce{CO} freeze-out stage of pre-stellar cores. This is in line with both ice and gas-phase observations of \ce{OCS} in protostars.
\end{itemize}

These findings suggest that the $\ce{CO} + \ce{SH}$ reaction is potentially responsible for a non-negligible share of the interstellar \ce{OCS} ice and therefore could be a valuable new addition to gas-grain chemical models focused on sulfur species. In turn, such models could help to disentangle the contributions of the $\ce{CO} + \ce{SH}$ and $\ce{CO} + \ce{S}$ routes to forming \ce{OCS}.
\begin{acknowledgements}
This work has been supported by the Danish National Research Foundation through the Center of Excellence “InterCat” (Grant agreement no.: DNRF150); the Netherlands Research School for Astronomy (NOVA); and the Dutch Astrochemistry Network II (DANII). KJC is grateful for support from NWO via a VENI fellowship (VI.Veni.212.296).
\end{acknowledgements}

%
%

   \bibliographystyle{aa} 
   \bibliography{mybib.bib} 

\begin{thebibliography}{93}
\expandafter\ifx\csname natexlab\endcsname\relax\def\natexlab#1{#1}\fi

\bibitem[{Adriaens {et~al.}(2010)Adriaens, Goumans, Catlow, \&
  Brown}]{Adriaens2010}
Adriaens, D., Goumans, T., Catlow, C., \& Brown, W. 2010, The Journal of
  Physical Chemistry C, 114, 1892

\bibitem[{{Altwegg} {et~al.}(2022){Altwegg}, {Combi}, {Fuselier}, {H{\"a}nni},
  {De Keyser}, {Mahjoub}, {M{\"u}ller}, {Pestoni}, {Rubin}, \&
  {Wampfler}}]{Altwegg2022}
{Altwegg}, K., {Combi}, M., {Fuselier}, S.~A., {et~al.} 2022, \mnras, 516, 3900

\bibitem[{{Artur de la Villarmois} {et~al.}(2023){Artur de la Villarmois},
  {Guzm{\'a}n}, {Yang}, {Zhang}, \& {Sakai}}]{delaVillarmois2023}
{Artur de la Villarmois}, E., {Guzm{\'a}n}, V.~V., {Yang}, Y.~L., {Zhang}, Y.,
  \& {Sakai}, N. 2023, \aap, 678, A124

\bibitem[{{Biver} {et~al.}(2021{\natexlab{a}}){Biver}, {Bockel{\'e}e-Morvan},
  {Boissier}, {Moreno}, {Crovisier}, {Lis}, {Colom}, {Cordiner}, {Milam},
  {Roth}, {Bonev}, {Dello Russo}, {Vervack}, \& {DiSanti}}]{Biver2021a}
{Biver}, N., {Bockel{\'e}e-Morvan}, D., {Boissier}, J., {et~al.}
  2021{\natexlab{a}}, \aap, 648, A49

\bibitem[{{Biver} {et~al.}(2021{\natexlab{b}}){Biver}, {Bockel{\'e}e-Morvan},
  {Lis}, {Despois}, {Moreno}, {Crovisier}, {Colom}, {Boissier}, {Dello Russo},
  {Vervack}, {Cordiner}, {Milam}, {Roth}, {Bonev}, {DiSanti}, {Davies}, \&
  {Kawakita}}]{Biver2021b}
{Biver}, N., {Bockel{\'e}e-Morvan}, D., {Lis}, D.~C., {et~al.}
  2021{\natexlab{b}}, \aap, 651, A25

\bibitem[{{Blake} {et~al.}(1987){Blake}, {Sutton}, {Masson}, \&
  {Phillips}}]{Blake1987}
{Blake}, G.~A., {Sutton}, E.~C., {Masson}, C.~R., \& {Phillips}, T.~G. 1987,
  \apj, 315, 621

\bibitem[{{Blake} {et~al.}(1994){Blake}, {van Dishoeck}, {Jansen}, {Groesbeck},
  \& {Mundy}}]{Blake1994}
{Blake}, G.~A., {van Dishoeck}, E.~F., {Jansen}, D.~J., {Groesbeck}, T.~D., \&
  {Mundy}, L.~G. 1994, \apj, 428, 680

\bibitem[{{Bockel{\'e}e-Morvan} {et~al.}(2000){Bockel{\'e}e-Morvan}, {Lis},
  {Wink}, {Despois}, {Crovisier}, {Bachiller}, {Benford}, {Biver}, {Colom},
  {Davies}, {G{\'e}rard}, {Germain}, {Houde}, {Mehringer}, {Moreno}, {Paubert},
  {Phillips}, \& {Rauer}}]{Bockelee-Morvan2000}
{Bockel{\'e}e-Morvan}, D., {Lis}, D.~C., {Wink}, J.~E., {et~al.} 2000, \aap,
  353, 1101

\bibitem[{{Boogert} {et~al.}(2022){Boogert}, {Brewer}, {Brittain}, \&
  {Emerson}}]{Boogert2022}
{Boogert}, A.~C.~A., {Brewer}, K., {Brittain}, A., \& {Emerson}, K.~S. 2022,
  \apj, 941, 32

\bibitem[{{Boogert} {et~al.}(2015){Boogert}, {Gerakines}, \&
  {Whittet}}]{Boogert2015}
{Boogert}, A.~C.~A., {Gerakines}, P.~A., \& {Whittet}, D. C.~B. 2015, \araa,
  53, 541

\bibitem[{{Boogert} {et~al.}(1997){Boogert}, {Schutte}, {Helmich}, {Tielens},
  \& {Wooden}}]{Boogert1997}
{Boogert}, A.~C.~A., {Schutte}, W.~A., {Helmich}, F.~P., {Tielens},
  A.~G.~G.~M., \& {Wooden}, D.~H. 1997, \aap, 317, 929

\bibitem[{Booth {et~al.}(2024)Booth, Temmink, van Dishoeck, Evans, Ilee, Kama,
  Keyte, Law, Leemker, van~der Marel, Nomura, Notsu, Öberg, \&
  Walsh}]{Booth2024}
Booth, A.~S., Temmink, M., van Dishoeck, E.~F., {et~al.} 2024, \aj
  [\eprint[arXiv]{2402.04002}]

\bibitem[{{Calmonte} {et~al.}(2016){Calmonte}, {Altwegg}, {Balsiger},
  {Berthelier}, {Bieler}, {Cessateur}, {Dhooghe}, {van Dishoeck}, {Fiethe},
  {Fuselier}, {Gasc}, {Gombosi}, {H{\"a}ssig}, {Le Roy}, {Rubin}, {S{\'e}mon},
  {Tzou}, \& {Wampfler}}]{Calmonte2016}
{Calmonte}, U., {Altwegg}, K., {Balsiger}, H., {et~al.} 2016, \mnras, 462, S253

\bibitem[{{Cartwright} {et~al.}(2020){Cartwright}, {Nordheim}, {Cruikshank},
  {Hand}, {Roser}, {Grundy}, {Beddingfield}, \& {Emery}}]{Cartwright2020}
{Cartwright}, R.~J., {Nordheim}, T.~A., {Cruikshank}, D.~P., {et~al.} 2020,
  \apjl, 902, L38

\bibitem[{{Cazaux} {et~al.}(2022){Cazaux}, {Carrascosa}, {Mu{\~n}oz Caro},
  {Caselli}, {Fuente}, {Navarro-Almaida}, \&
  {Rivi{\'e}re-Marichalar}}]{Cazaux2022}
{Cazaux}, S., {Carrascosa}, H., {Mu{\~n}oz Caro}, G.~M., {et~al.} 2022, \aap,
  657, A100

\bibitem[{{Ceccarelli} {et~al.}(2014){Ceccarelli}, {Caselli},
  {Bockel{\'e}e-Morvan}, {Mousis}, {Pizzarello}, {Robert}, \&
  {Semenov}}]{Ceccarelli2014}
{Ceccarelli}, C., {Caselli}, P., {Bockel{\'e}e-Morvan}, D., {et~al.} 2014, in
  Protostars and Planets VI, ed. H.~{Beuther}, R.~S. {Klessen}, C.~P.
  {Dullemond}, \& T.~{Henning}, 859--882

\bibitem[{{Cernicharo} {et~al.}(2012){Cernicharo}, {Marcelino}, {Roueff},
  {Gerin}, {Jim{\'e}nez-Escobar}, \& {Mu{\~n}oz Caro}}]{Cernicharo2012}
{Cernicharo}, J., {Marcelino}, N., {Roueff}, E., {et~al.} 2012, \apjl, 759, L43

\bibitem[{{Charnley} {et~al.}(1992){Charnley}, {Tielens}, \&
  {Millar}}]{Charnley1992}
{Charnley}, S.~B., {Tielens}, A.~G.~G.~M., \& {Millar}, T.~J. 1992, \apjl, 399,
  L71

\bibitem[{{Chen} {et~al.}(2015){Chen}, {Juang}, {Nuevo}, {Jim{\'e}nez-Escobar},
  {Mu{\~n}oz Caro}, {Qiu}, {Chu}, {Yih}, {Wu}, {Fung}, \& {Ip}}]{Chen2015}
{Chen}, Y.~J., {Juang}, K.~J., {Nuevo}, M., {et~al.} 2015, \apj, 798, 80

\bibitem[{{Codella} {et~al.}(2021){Codella}, {Bianchi}, {Podio}, {Mercimek},
  {Ceccarelli}, {L{\'o}pez-Sepulcre}, {Bachiller}, {Caselli}, {Sakai}, {Neri},
  {Fontani}, {Favre}, {Balucani}, {Lefloch}, {Viti}, \&
  {Yamamoto}}]{Codella2021}
{Codella}, C., {Bianchi}, E., {Podio}, L., {et~al.} 2021, \aap, 654, A52

\bibitem[{{Drdla} {et~al.}(1989){Drdla}, {Knapp}, \& {van
  Dishoeck}}]{Drdla1989}
{Drdla}, K., {Knapp}, G.~R., \& {van Dishoeck}, E.~F. 1989, \apj, 345, 815

\bibitem[{{Drozdovskaya} {et~al.}(2018){Drozdovskaya}, {van Dishoeck},
  {J{\o}rgensen}, {Calmonte}, {van der Wiel}, {Coutens}, {Calcutt},
  {M{\"u}ller}, {Bjerkeli}, {Persson}, {Wampfler}, \&
  {Altwegg}}]{Drozdovskaya2018}
{Drozdovskaya}, M.~N., {van Dishoeck}, E.~F., {J{\o}rgensen}, J.~K., {et~al.}
  2018, \mnras, 476, 4949

\bibitem[{{Druard} \& {Wakelam}(2012)}]{Druard2012}
{Druard}, C. \& {Wakelam}, V. 2012, \mnras, 426, 354

\bibitem[{{Esplugues} {et~al.}(2022){Esplugues}, {Fuente}, {Navarro-Almaida},
  {Rodr{\'\i}guez-Baras}, {Majumdar}, {Caselli}, {Wakelam}, {Roueff},
  {Bachiller}, {Spezzano}, {Rivi{\`e}re-Marichalar},
  {Mart{\'\i}n-Dom{\'e}nech}, \& {Mu{\~n}oz Caro}}]{Esplugues2022}
{Esplugues}, G., {Fuente}, A., {Navarro-Almaida}, D., {et~al.} 2022, \aap, 662,
  A52

\bibitem[{{Esplugues} {et~al.}(2014){Esplugues}, {Viti}, {Goicoechea}, \&
  {Cernicharo}}]{Esplugues2014}
{Esplugues}, G.~B., {Viti}, S., {Goicoechea}, J.~R., \& {Cernicharo}, J. 2014,
  \aap, 567, A95

\bibitem[{{Fedoseev} {et~al.}(2015){Fedoseev}, {Ioppolo}, \&
  {Linnartz}}]{Fedoseev2015}
{Fedoseev}, G., {Ioppolo}, S., \& {Linnartz}, H. 2015, \mnras, 446, 449

\bibitem[{{Ferrante} {et~al.}(2008){Ferrante}, {Moore}, {Spiliotis}, \&
  {Hudson}}]{Ferrante2008}
{Ferrante}, R.~F., {Moore}, M.~H., {Spiliotis}, M.~M., \& {Hudson}, R.~L. 2008,
  \apj, 684, 1210

\bibitem[{{Fontani} {et~al.}(2023){Fontani}, {Roueff}, {Colzi}, \&
  {Caselli}}]{Fontani2023}
{Fontani}, F., {Roueff}, E., {Colzi}, L., \& {Caselli}, P. 2023, \aap, 680, A58

\bibitem[{{Fuchs} {et~al.}(2009){Fuchs}, {Cuppen}, {Ioppolo}, {Romanzin},
  {Bisschop}, {Andersson}, {van Dishoeck}, \& {Linnartz}}]{Fuchs2009}
{Fuchs}, G.~W., {Cuppen}, H.~M., {Ioppolo}, S., {et~al.} 2009, \aap, 505, 629

\bibitem[{{Fuente} {et~al.}(2010){Fuente}, {Cernicharo}, {Ag{\'u}ndez},
  {Bern{\'e}}, {Goicoechea}, {Alonso-Albi}, \& {Marcelino}}]{Fuente2010}
{Fuente}, A., {Cernicharo}, J., {Ag{\'u}ndez}, M., {et~al.} 2010, \aap, 524,
  A19

\bibitem[{{Garozzo} {et~al.}(2010){Garozzo}, {Fulvio}, {Kanuchova}, {Palumbo},
  \& {Strazzulla}}]{Garozzo2010}
{Garozzo}, M., {Fulvio}, D., {Kanuchova}, Z., {Palumbo}, M.~E., \&
  {Strazzulla}, G. 2010, \aap, 509, A67

\bibitem[{{Garrod} {et~al.}(2007){Garrod}, {Wakelam}, \& {Herbst}}]{Garrod2007}
{Garrod}, R.~T., {Wakelam}, V., \& {Herbst}, E. 2007, \aap, 467, 1103

\bibitem[{{Gibb} {et~al.}(2000){Gibb}, {Nummelin}, {Irvine}, {Whittet}, \&
  {Bergman}}]{Gibb2000}
{Gibb}, E., {Nummelin}, A., {Irvine}, W.~M., {Whittet}, D.~C.~B., \& {Bergman},
  P. 2000, \apj, 545, 309

\bibitem[{{Goldsmith} \& {Li}(2005)}]{Goldsmith2005}
{Goldsmith}, P.~F. \& {Li}, D. 2005, \apj, 622, 938

\bibitem[{{Herpin} {et~al.}(2009){Herpin}, {Marseille}, {Wakelam}, {Bontemps},
  \& {Lis}}]{Herpin2009}
{Herpin}, F., {Marseille}, M., {Wakelam}, V., {Bontemps}, S., \& {Lis}, D.~C.
  2009, \aap, 504, 853

\bibitem[{{Hibbitts} {et~al.}(2000){Hibbitts}, {McCord}, \&
  {Hansen}}]{Hibbitts2000}
{Hibbitts}, C.~A., {McCord}, T.~B., \& {Hansen}, G.~B. 2000, \jgr, 105, 22541

\bibitem[{{Hiraoka} {et~al.}(1994){Hiraoka}, {Ohashi}, {Kihara}, {Yamamoto},
  {Sato}, \& {Yamashita}}]{Hiraoka1994}
{Hiraoka}, K., {Ohashi}, N., {Kihara}, Y., {et~al.} 1994, Chemical Physics
  Letters, 229, 408

\bibitem[{{Hudgins} {et~al.}(1993){Hudgins}, {Sandford}, {Allamandola}, \&
  {Tielens}}]{Hudgins1993}
{Hudgins}, D.~M., {Sandford}, S.~A., {Allamandola}, L.~J., \& {Tielens},
  A.~G.~G.~M. 1993, \apjs, 86, 713

\bibitem[{{Ioppolo} {et~al.}(2013){Ioppolo}, {Fedoseev}, {Lamberts},
  {Romanzin}, \& {Linnartz}}]{Ioppolo2013}
{Ioppolo}, S., {Fedoseev}, G., {Lamberts}, T., {Romanzin}, C., \& {Linnartz},
  H. 2013, Rev. Sci. Instrum., 84, 073112

\bibitem[{{Ioppolo} {et~al.}(2011){Ioppolo}, {van Boheemen}, {Cuppen}, {van
  Dishoeck}, \& {Linnartz}}]{Ioppolo2011}
{Ioppolo}, S., {van Boheemen}, Y., {Cuppen}, H.~M., {van Dishoeck}, E.~F., \&
  {Linnartz}, H. 2011, \mnras, 413, 2281

\bibitem[{{Jessup} {et~al.}(2007){Jessup}, {Spencer}, \& {Yelle}}]{Jessup2007}
{Jessup}, K.~L., {Spencer}, J., \& {Yelle}, R. 2007, \icarus, 192, 24

\bibitem[{{Jim{\'e}nez-Escobar} \& {Mu{\~n}oz Caro}(2011)}]{JimenezEscobar2011}
{Jim{\'e}nez-Escobar}, A. \& {Mu{\~n}oz Caro}, G.~M. 2011, Astronomy and
  Astrophysics, 536, A91

\bibitem[{{Jim{\'e}nez-Escobar} {et~al.}(2014){Jim{\'e}nez-Escobar}, {Mu{\~n}oz
  Caro}, \& {Chen}}]{Jimenez-Escobar2014}
{Jim{\'e}nez-Escobar}, A., {Mu{\~n}oz Caro}, G.~M., \& {Chen}, Y.~J. 2014,
  \mnras, 443, 343

\bibitem[{{Kolesnikov{\'a}} {et~al.}(2014){Kolesnikov{\'a}}, {Tercero},
  {Cernicharo}, {Alonso}, {Daly}, {Gordon}, \& {Shipman}}]{Koleniskova2014}
{Kolesnikov{\'a}}, L., {Tercero}, B., {Cernicharo}, J., {et~al.} 2014, \apjl,
  784, L7

\bibitem[{{Kushwahaa} {et~al.}(2023){Kushwahaa}, {Drozdovskaya}, {Tychoniec},
  \& {Tabone}}]{Kushwahaa2023}
{Kushwahaa}, T., {Drozdovskaya}, M.~N., {Tychoniec}, {\L}., \& {Tabone}, B.
  2023, \aap, 672, A122

\bibitem[{{Laas} \& {Caselli}(2019)}]{Laas2019}
{Laas}, J.~C. \& {Caselli}, P. 2019, \aap, 624, A108

\bibitem[{{Lacy} {et~al.}(1994){Lacy}, {Knacke}, {Geballe}, \&
  {Tokunaga}}]{Lacy1994}
{Lacy}, J.~H., {Knacke}, R., {Geballe}, T.~R., \& {Tokunaga}, A.~T. 1994,
  \apjl, 428, L69

\bibitem[{Lamberts \& Kästner(2017)}]{Lamberts2017}
Lamberts, T. \& Kästner, J. 2017, The Journal of Physical Chemistry A, 121,
  9736, pMID: 29190103

\bibitem[{{Le Gal} {et~al.}(2019){Le Gal}, {{\"O}berg}, {Loomis}, {Pegues}, \&
  {Bergner}}]{LeGal2019}
{Le Gal}, R., {{\"O}berg}, K.~I., {Loomis}, R.~A., {Pegues}, J., \& {Bergner},
  J.~B. 2019, \apj, 876, 72

\bibitem[{{Le Gal} {et~al.}(2021){Le Gal}, {{\"O}berg}, {Teague}, {Loomis},
  {Law}, {Walsh}, {Bergin}, {M{\'e}nard}, {Wilner}, {Andrews}, {Aikawa},
  {Booth}, {Cataldi}, {Bergner}, {Bosman}, {Cleeves}, {Czekala}, {Furuya},
  {Guzm{\'a}n}, {Huang}, {Ilee}, {Nomura}, {Qi}, {Schwarz}, {Tsukagoshi},
  {Yamato}, \& {Zhang}}]{LeGal2021}
{Le Gal}, R., {{\"O}berg}, K.~I., {Teague}, R., {et~al.} 2021, \apjs, 257, 12

\bibitem[{{Li} {et~al.}(2015){Li}, {Wang}, {Zhu}, {Zhang}, \& {Li}}]{Li2015}
{Li}, J., {Wang}, J., {Zhu}, Q., {Zhang}, J., \& {Li}, D. 2015, \apj, 802, 40

\bibitem[{{Linke} {et~al.}(1979){Linke}, {Frerking}, \& {Thaddeus}}]{Linke1979}
{Linke}, R.~A., {Frerking}, M.~A., \& {Thaddeus}, P. 1979, \apjl, 234, L139

\bibitem[{{Loison} {et~al.}(2012){Loison}, {Halvick}, {Bergeat}, {Hickson}, \&
  {Wakelam}}]{Loison2012}
{Loison}, J.-C., {Halvick}, P., {Bergeat}, A., {Hickson}, K.~M., \& {Wakelam},
  V. 2012, \mnras, 421, 1476

\bibitem[{{L{\'o}pez-Gallifa} {et~al.}(2024){L{\'o}pez-Gallifa}, {Rivilla},
  {Beltr{\'a}n}, {Colzi}, {Mininni}, {S{\'a}nchez-Monge}, {Fontani}, {Viti},
  {Jim{\'e}nez-Serra}, {Testi}, {Cesaroni}, \& {Lorenzani}}]{Lopez-Gallifa2024}
{L{\'o}pez-Gallifa}, {\'A}., {Rivilla}, V.~M., {Beltr{\'a}n}, M.~T., {et~al.}
  2024, \mnras, 529, 3244

\bibitem[{{Majumdar} {et~al.}(2016){Majumdar}, {Gratier}, {Vidal}, {Wakelam},
  {Loison}, {Hickson}, \& {Caux}}]{Majumdar2016}
{Majumdar}, L., {Gratier}, P., {Vidal}, T., {et~al.} 2016, \mnras, 458, 1859

\bibitem[{McAnally {et~al.}(2024)McAnally, Bockov{\'a}, Herath, Turner,
  Meinert, \& Kaiser}]{Mcanally2024}
McAnally, M., Bockov{\'a}, J., Herath, A., {et~al.} 2024, Nature
  Communications, 15, 4409

\bibitem[{{McClure} {et~al.}(2023){McClure}, {Rocha}, {Pontoppidan}, {Crouzet},
  {Chu}, {Dartois}, {Lamberts}, {Noble}, {Pendleton}, {Perotti}, {Qasim},
  {Rachid}, {Smith}, {Sun}, {Beck}, {Boogert}, {Brown}, {Caselli}, {Charnley},
  {Cuppen}, {Dickinson}, {Drozdovskaya}, {Egami}, {Erkal}, {Fraser}, {Garrod},
  {Harsono}, {Ioppolo}, {Jim{\'e}nez-Serra}, {Jin}, {J{\o}rgensen},
  {Kristensen}, {Lis}, {McCoustra}, {McGuire}, {Melnick}, {{\~A}-berg},
  {Palumbo}, {Shimonishi}, {Sturm}, {van Dishoeck}, \&
  {Linnartz}}]{McClure2023}
{McClure}, M.~K., {Rocha}, W.~R.~M., {Pontoppidan}, K.~M., {et~al.} 2023,
  Nature Astronomy, 7, 431

\bibitem[{{Molpeceres} {et~al.}(2023){Molpeceres}, {Enrique-Romero}, \&
  {Aikawa}}]{Molpeceres2023}
{Molpeceres}, G., {Enrique-Romero}, J., \& {Aikawa}, Y. 2023, \aap, 677, A39

\bibitem[{{Molpeceres} {et~al.}(2021){Molpeceres}, {Garc{\'\i}a de la
  Concepci{\'o}n}, \& {Jim{\'e}nez-Serra}}]{Molpeceres2021}
{Molpeceres}, G., {Garc{\'\i}a de la Concepci{\'o}n}, J., \&
  {Jim{\'e}nez-Serra}, I. 2021, \apj, 923, 159

\bibitem[{{Moullet} {et~al.}(2013){Moullet}, {Lellouch}, {Moreno}, {Gurwell},
  {Black}, \& {Butler}}]{Moullet2013}
{Moullet}, A., {Lellouch}, E., {Moreno}, R., {et~al.} 2013, \apj, 776, 32

\bibitem[{{Moullet} {et~al.}(2008){Moullet}, {Lellouch}, {Moreno}, {Gurwell},
  \& {Moore}}]{Moullet2008}
{Moullet}, A., {Lellouch}, E., {Moreno}, R., {Gurwell}, M.~A., \& {Moore}, C.
  2008, \aap, 482, 279

\bibitem[{{M{\"u}ller} {et~al.}(2016){M{\"u}ller}, {Belloche}, {Xu}, {Lees},
  {Garrod}, {Walters}, {van Wijngaarden}, {Lewen}, {Schlemmer}, \&
  {Menten}}]{Muller2016}
{M{\"u}ller}, H. S.~P., {Belloche}, A., {Xu}, L.-H., {et~al.} 2016, \aap, 587,
  A92

\bibitem[{{Navarro-Almaida} {et~al.}(2020){Navarro-Almaida}, {Le Gal},
  {Fuente}, {Rivi{\`e}re-Marichalar}, {Wakelam}, {Cazaux}, {Caselli}, {Laas},
  {Alonso-Albi}, {Loison}, {Gerin}, {Kramer}, {Roueff}, {Bachiller},
  {Commer{\c{c}}on}, {Friesen}, {Garc{\'\i}a-Burillo}, {Goicoechea},
  {Giuliano}, {Jim{\'e}nez-Serra}, {Kirk}, {Lattanzi}, {Malinen}, {Marcelino},
  {Mart{\'\i}n-Dom{\`e}nech}, {Mu{\~n}oz Caro}, {Pineda}, {Tercero},
  {Trevi{\~n}o-Morales}, {Roncero}, {Hacar}, {Tafalla}, \&
  {Ward-Thompson}}]{Navarro-Almaida2020}
{Navarro-Almaida}, D., {Le Gal}, R., {Fuente}, A., {et~al.} 2020, \aap, 637,
  A39

\bibitem[{{Nguyen} {et~al.}(2021){Nguyen}, {Oba}, {Sameera}, {Kouchi}, \&
  {Watanabe}}]{Nguyen2021}
{Nguyen}, T., {Oba}, Y., {Sameera}, W.~M.~C., {Kouchi}, A., \& {Watanabe}, N.
  2021, \apj, 922, 146

\bibitem[{{Noble} {et~al.}(2011){Noble}, {Dulieu}, {Congiu}, \&
  {Fraser}}]{Noble2011}
{Noble}, J.~A., {Dulieu}, F., {Congiu}, E., \& {Fraser}, H.~J. 2011, \apj, 735,
  121

\bibitem[{{Oba} {et~al.}(2019){Oba}, {Tomaru}, {Kouchi}, \&
  {Watanabe}}]{Oba2019}
{Oba}, Y., {Tomaru}, T., {Kouchi}, A., \& {Watanabe}, N. 2019, \apj, 874, 124

\bibitem[{{Oba} {et~al.}(2018){Oba}, {Tomaru}, {Lamberts}, {Kouchi}, \&
  {Watanabe}}]{Oba2018}
{Oba}, Y., {Tomaru}, T., {Lamberts}, T., {Kouchi}, A., \& {Watanabe}, N. 2018,
  Nature Astronomy, 2, 228

\bibitem[{{Oba} {et~al.}(2010){Oba}, {Watanabe}, {Kouchi}, {Hama}, \&
  {Pirronello}}]{Oba2010a}
{Oba}, Y., {Watanabe}, N., {Kouchi}, A., {Hama}, T., \& {Pirronello}, V. 2010,
  \apjl, 712, L174

\bibitem[{{{\"O}berg} {et~al.}(2008){{\"O}berg}, {Boogert}, {Pontoppidan},
  {Blake}, {Evans}, {Lahuis}, \& {van Dishoeck}}]{Oberg2008}
{{\"O}berg}, K.~I., {Boogert}, A.~C.~A., {Pontoppidan}, K.~M., {et~al.} 2008,
  \apj, 678, 1032

\bibitem[{{Oya} {et~al.}(2016){Oya}, {Sakai}, {L{\'o}pez-Sepulcre}, {Watanabe},
  {Ceccarelli}, {Lefloch}, {Favre}, \& {Yamamoto}}]{Oya2016}
{Oya}, Y., {Sakai}, N., {L{\'o}pez-Sepulcre}, A., {et~al.} 2016, \apj, 824, 88

\bibitem[{{Palumbo} {et~al.}(1997){Palumbo}, {Geballe}, \&
  {Tielens}}]{Palumbo1997}
{Palumbo}, M.~E., {Geballe}, T.~R., \& {Tielens}, A.~G.~G.~M. 1997, \apj, 479,
  839

\bibitem[{{Palumbo} {et~al.}(1995){Palumbo}, {Tielens}, \&
  {Tokunaga}}]{Palumbo1995}
{Palumbo}, M.~E., {Tielens}, A.~G.~G.~M., \& {Tokunaga}, A.~T. 1995, \apj, 449,
  674

\bibitem[{{Phuong} {et~al.}(2018){Phuong}, {Chapillon}, {Majumdar}, {Dutrey},
  {Guilloteau}, {Pi{\'e}tu}, {Wakelam}, {Diep}, {Tang}, {Beck}, \&
  {Bary}}]{Phuong2018}
{Phuong}, N.~T., {Chapillon}, E., {Majumdar}, L., {et~al.} 2018, \aap, 616, L5

\bibitem[{{Qasim} {et~al.}(2019){Qasim}, {Lamberts}, {He}, {Chuang},
  {Fedoseev}, {Ioppolo}, {Boogert}, \& {Linnartz}}]{Qasim2019}
{Qasim}, D., {Lamberts}, T., {He}, J., {et~al.} 2019, \aap, 626, A118

\bibitem[{{Qasim} {et~al.}(2020){Qasim}, {Witlox}, {Fedoseev}, {Chuang},
  {Banu}, {Krasnokutski}, {Ioppolo}, {K{\"a}stner}, {van Dishoeck}, \&
  {Linnartz}}]{Qasim2020}
{Qasim}, D., {Witlox}, M.~J.~A., {Fedoseev}, G., {et~al.} 2020, Rev. Sci.
  Instrum., 91, 054501

\bibitem[{{Rivi{\`e}re-Marichalar} {et~al.}(2021){Rivi{\`e}re-Marichalar},
  {Fuente}, {Le Gal}, {Arabhavi}, {Cazaux}, {Navarro-Almaida}, {Ribas},
  {Mendigut{\'\i}a}, {Barrado}, \& {Montesinos}}]{Riviere-Marichalar2021}
{Rivi{\`e}re-Marichalar}, P., {Fuente}, A., {Le Gal}, R., {et~al.} 2021, \aap,
  652, A46

\bibitem[{{Rocha} {et~al.}(2024){Rocha}, {van Dishoeck}, {Ressler}, {van
  Gelder}, {Slavicinska}, {Brunken}, {Linnartz}, {Ray}, {Beuther}, {Caratti o
  Garatti}, {Geers}, {Kavanagh}, {Klaassen}, {Justtanont}, {Chen}, {Francis},
  {Gieser}, {Perotti}, {Tychoniec}, {Barsony}, {Majumdar}, {le Gouellec},
  {Chu}, {Lew}, {Henning}, \& {Wright}}]{Rocha2024}
{Rocha}, W.~R.~M., {van Dishoeck}, E.~F., {Ressler}, M.~E., {et~al.} 2024,
  \aap, 683, A124

\bibitem[{{Rodr{\'\i}guez-Almeida} {et~al.}(2021){Rodr{\'\i}guez-Almeida},
  {Jim{\'e}nez-Serra}, {Rivilla}, {Mart{\'\i}n-Pintado}, {Zeng}, {Tercero}, {de
  Vicente}, {Colzi}, {Rico-Villas}, {Mart{\'\i}n}, \&
  {Requena-Torres}}]{Rodriguez-Almeida2021}
{Rodr{\'\i}guez-Almeida}, L.~F., {Jim{\'e}nez-Serra}, I., {Rivilla}, V.~M.,
  {et~al.} 2021, \apjl, 912, L11

\bibitem[{{Santos} {et~al.}(2022){Santos}, {Chuang}, {Lamberts}, {Fedoseev},
  {Ioppolo}, \& {Linnartz}}]{Santos2022}
{Santos}, J.~C., {Chuang}, K.-J., {Lamberts}, T., {et~al.} 2022, \apjl, 931,
  L33

\bibitem[{{Santos} {et~al.}(2023){Santos}, {Linnartz}, \&
  {Chuang}}]{Santos2023}
{Santos}, J.~C., {Linnartz}, H., \& {Chuang}, K.~J. 2023, \aap, 678, A112

\bibitem[{{Santos} {et~al.}(submitted){Santos}, {van Gelder}, {Nazari},
  {Ahmadi}, \& {van Dishoeck}}]{Santos2024_submm}
{Santos}, J.~C., {van Gelder}, M.~L., {Nazari}, P., {Ahmadi}, A., \& {van
  Dishoeck}, E.~F. submitted, \aap

\bibitem[{{Semenov} {et~al.}(2018){Semenov}, {Favre}, {Fedele}, {Guilloteau},
  {Teague}, {Henning}, {Dutrey}, {Chapillon}, {Hersant}, \&
  {Pi{\'e}tu}}]{Semenov2018}
{Semenov}, D., {Favre}, C., {Fedele}, D., {et~al.} 2018, \aap, 617, A28

\bibitem[{{Shingledecker} {et~al.}(2020){Shingledecker}, {Lamberts}, {Laas},
  {Vasyunin}, {Herbst}, {K{\"a}stner}, \& {Caselli}}]{Shingledecker2020}
{Shingledecker}, C.~N., {Lamberts}, T., {Laas}, J.~C., {et~al.} 2020, \apj,
  888, 52

\bibitem[{{Smith} {et~al.}(1980){Smith}, {Stecher}, \& {Casswell}}]{Smith1980}
{Smith}, A.~M., {Stecher}, T.~P., \& {Casswell}, L. 1980, \apj, 242, 402

\bibitem[{{Spezzano} {et~al.}(2022){Spezzano}, {Sipil{\"a}}, {Caselli},
  {Jensen}, {Czakli}, {Bizzocchi}, {Chantzos}, {Esplugues}, {Fuente}, \&
  {Eisenhauer}}]{Spezzano2022}
{Spezzano}, S., {Sipil{\"a}}, O., {Caselli}, P., {et~al.} 2022, \aap, 661, A111

\bibitem[{{Tielens} \& {Hagen}(1982)}]{Tielens1982}
{Tielens}, A.~G.~G.~M. \& {Hagen}, W. 1982, \aap, 114, 245

\bibitem[{{Tschersich}(2000)}]{Tschersich2000}
{Tschersich}, K.~G. 2000, J. Appl. Phys., 87, 2565

\bibitem[{{van der Tak} {et~al.}(2003){van der Tak}, {Boonman}, {Braakman}, \&
  {van Dishoeck}}]{vanderTak2003}
{van der Tak}, F.~F.~S., {Boonman}, A.~M.~S., {Braakman}, R., \& {van
  Dishoeck}, E.~F. 2003, \aap, 412, 133

\bibitem[{{Vidal} {et~al.}(2017){Vidal}, {Loison}, {Jaziri}, {Ruaud},
  {Gratier}, \& {Wakelam}}]{Vidal2017}
{Vidal}, T. H.~G., {Loison}, J.-C., {Jaziri}, A.~Y., {et~al.} 2017, \mnras,
  469, 435

\bibitem[{Wang {et~al.}(2022)Wang, Marks, Tuli, Mebel, Azyazov, \&
  Kaiser}]{Wang2022}
Wang, J., Marks, J.~H., Tuli, L.~B., {et~al.} 2022, The Journal of Physical
  Chemistry A, 126, 9699

\bibitem[{{Watanabe} \& {Kouchi}(2002)}]{Watanabe2002}
{Watanabe}, N. \& {Kouchi}, A. 2002, \apjl, 571, L173

\bibitem[{{Yarnall} \& {Hudson}(2022)}]{Yarnall2022}
{Yarnall}, Y.~Y. \& {Hudson}, R.~L. 2022, \apjl, 931, L4

\bibitem[{{Zapata} {et~al.}(2015){Zapata}, {Palau}, {Galv{\'a}n-Madrid},
  {Rodr{\'\i}guez}, {Garay}, {Moran}, \& {Franco-Hern{\'a}ndez}}]{Zapata2015}
{Zapata}, L.~A., {Palau}, A., {Galv{\'a}n-Madrid}, R., {et~al.} 2015, \mnras,
  447, 1826

\end{thebibliography}

\end{document}